\documentclass[twocolumn,fleqn,10pt]{wlscirep}
\usepackage[utf8]{inputenc}
\usepackage[T1]{fontenc}
\usepackage{float}
\usepackage{pgfplots}
\usepackage{lineno}
\usepackage{setspace}

\newcommand{\bibcommenthead}[1]{} 

\title{Correlation between Gyral Size, Brain Size, and Head Impact Risk across Mammalian Species}
\author[1]{Nianqin Zhang}
\author[2]{Yongjun Zhang}
\affil[1]{Department of Cardiology, Fuwai Hospital, National Center for Cardiovascular Diseases, Chinese Academy of Medical Sciences and Peking Union Medical College, Beijing, 100037, China}
\affil[2]{Science College, Liaoning Technical University, Fuxin, 123000, China}

\begin{abstract}
A study on primates has established that gyral size is largely independent of overall brain size. Building on this—and other research suggesting that brain gyrification may mitigate the effects of head impacts—our study aims to explore potential correlations between gyral size and the risk of head impact across a diverse range of mammalian species. Our findings corroborate the idea that gyral sizes are largely independent of brain sizes, especially among species with larger brains, thus extending this observation beyond primates. Preliminary evidence also suggests a correlation between an animal's gyral size and its lifestyle, particularly in terms of head-impact risk. For instance, goats, known for their headbutting behaviors, exhibit smaller gyral sizes. In contrast, species such as manatees and dugongs, which typically face lower risks of head impact, have lissencephalic brains. Additionally, we explore mechanisms that may explain how narrower gyral sizes could offer protective advantages against head impact. Finally, we discuss a possible trade-off associated with gyrencephaly.

\noindent \textbf{Keywords:} Evolution, Mammals, Neuroanatomy, Phylogenesis, Acoustic Waves
\end{abstract}

\begin{document}

\maketitle

\section{Introduction}
The majority of large mammals feature gyrencephalic brain structures \cite{www.brainmuseum.org}, defined by the presence of cortical folds known as gyri. These gyri are separated by sulci, which are grooves that extend into the brain tissue, thus increasing the cortical surface area and potentially enabling a higher neuron count \cite{10.1007/978-1-4615-3824-0_1}. However, gyrification—the degree to which the cortex is folded—varies significantly among species. For example, ungulates generally demonstrate more complex patterns of gyrification compared to primates \cite{10.1111/j.1460-9568.2007.05524.x}.

Extensive research has delved into the mechanisms underlying cortical folding \cite{10.1146/annurev-neuro-071714-034128,10.1016/j.pneurobio.2021.102111}, with studies focusing on both developmental factors \cite{10.1093/cercor/bhi068,10.1038/nrn2008} and biomechanical influences \cite{10.1126/science.1135626,10.1073/pnas.2016830117,10.1016/j.actbio.2021.07.044,10.1016/j.jbiomech.2021.110851,10.1002/jez.1401130304}. Despite these efforts, the functional role of cortical folding remains an open question. In a computational study by Sáez et al. \cite{10.1016/j.ijmecsci.2020.105914}, it was found that geometric factors, such as the Gyrification Index (GI), significantly influence the brain's mechanical response to impacts. These findings suggest that gyrification serves as a damping mechanism to mitigate mechanical trauma, particularly in larger-brained mammals. However, since the GI, calculated as the ratio of the total surface area of the brain to its convex hull \cite{10.1007/bf00304699,10.1016/j.tins.2013.01.006}, is a dimensionless metric. Given that different species exhibit similar brain tissue properties \cite{10.1016/j.neuroimage.2010.03.077}, there is a crucial need for a length-based parameter for more comprehensive cross-species comparisons.

Heuer et al. \cite{10.1016/j.cortex.2019.04.011} observed a consistent gyral size or "fold wavelength" of approximately 12 mm across various gyrencephalic primate neocortices, despite differing cerebral volumes. This uniformity could be attributed to similar neocortical stiffness across these species but might also reflect a universal need for mitigating head impact risks. Such impacts can induce varying acceleration levels in brain tissue, causing shearing stress and the potential for diffuse axonal injury \cite{Gennarelli1987}. Moreover, these impacts can generate acoustical waves that are harmful to brain tissue.

This study aims to corroborate the idea that gyral sizes are not strongly correlated with brain sizes. We also explore a potential link between gyral size and an animal's lifestyle, especially concerning head impact risks. Furthermore, we explore two possible mechanisms through which narrower gyral sizes could provide protective advantages against head impacts.

\section{Methods}
\subsection{Quantifying Gyral Size in Gyrencephalic Brains}

Understanding gyral size is pivotal for comparative analyses of gyrification patterns across various species. Two methods are employed for its quantification.

\subsubsection{Characterization by Width}
\begin{figure}[htbp]
\centering
\includegraphics[height=3.6cm]{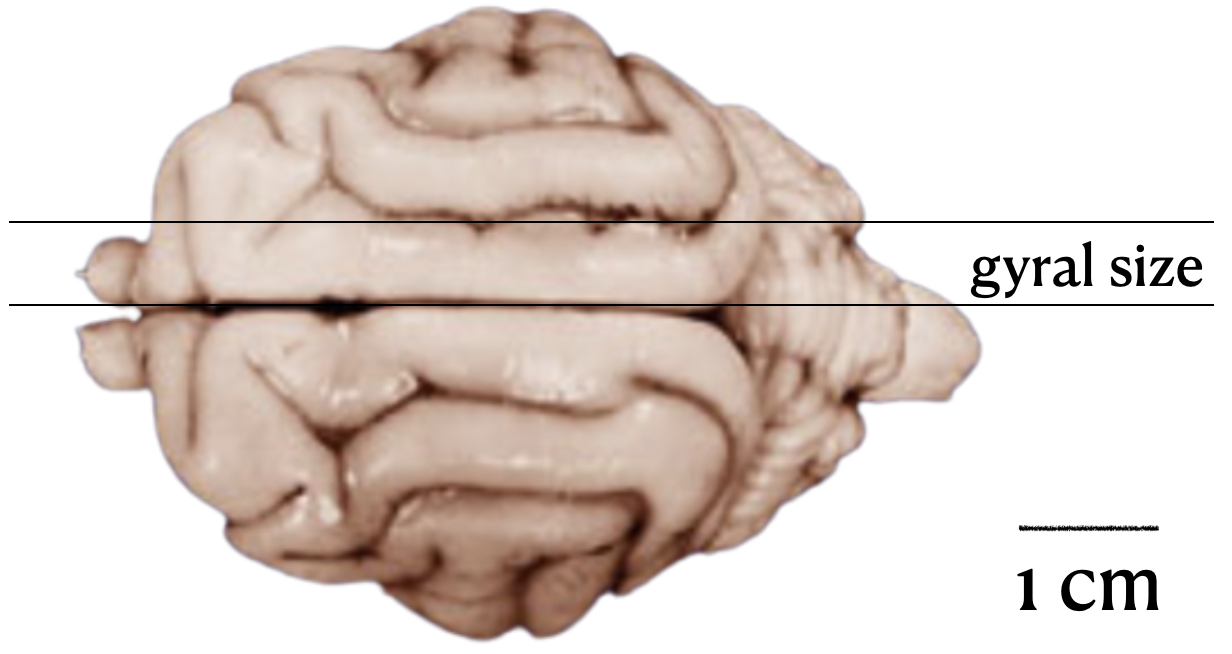}
\caption{
Gyral size defined through characteristic width, exemplified using a cat's brain \cite{www.brainmuseum.org}. The image, as well as other brain images for demonstration purposes throughout this paper, are sourced from the Mammalian Brain Collections at www.brainmuseum.org, property of the University of Wisconsin and the Michigan State Comparative Mammalian Brain Collections, and funded by the National Science Foundation and the National Institutes of Health. 
}
\label{GS1}
\end{figure}

The first approach defines gyral size as the characteristic width of gyri, as illustrated in Fig.~\ref{GS1}. This method aligns with Prothero and Sundsten's gyral width concept \cite{10.1159/000121313} and is consistent with the fold wavelength used by Heuer et al. \cite{10.1016/j.cortex.2019.04.011}.

\subsubsection{Computational Method}
\begin{figure}[htbp]
\centering
\includegraphics[height=3.6cm]{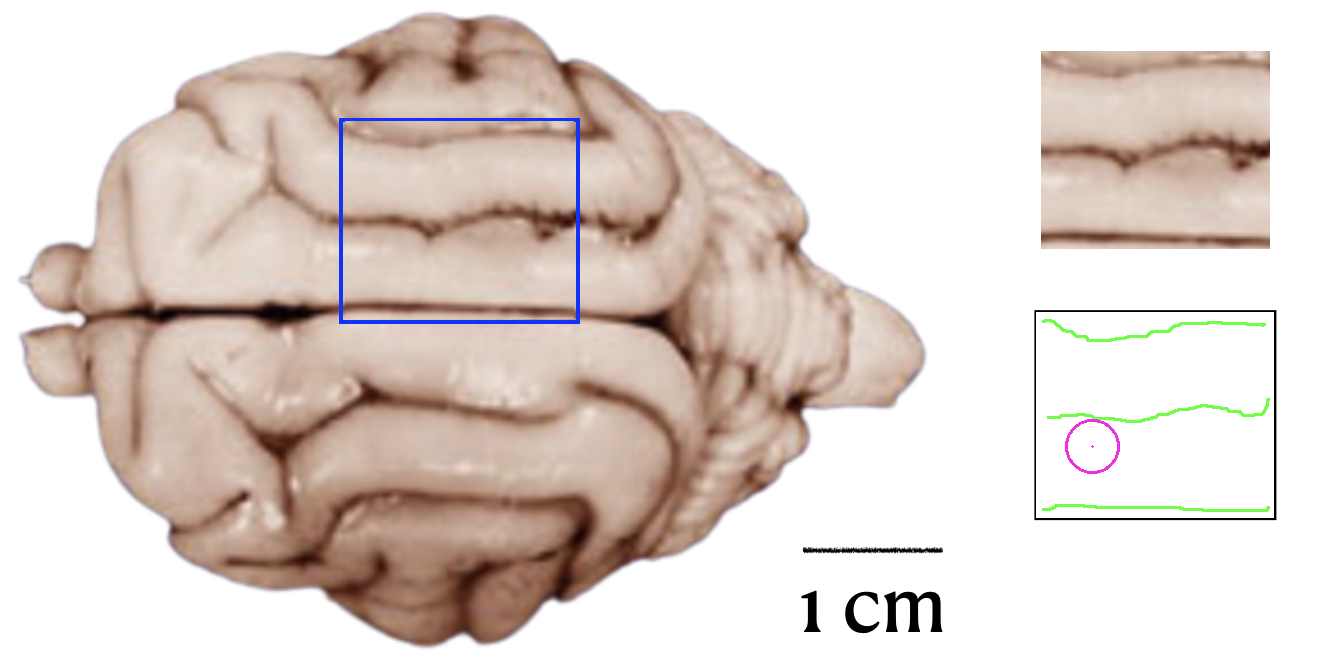}
\caption{
Computational approach for gyral size estimation. This method identifies the positions of sulci on the surface image of the brain and selects a random starting point. From this point, a circle is expanded until it encounters the closest sulcus. Multiple iterations estimate an average radius, quadrupled to derive gyral size. Code example available at \href{https://github.com/zyjlntu/brain_gyral_size}{https://github.com/zyjlntu/brain\_gyral\_size}.}
\label{GS2}
\end{figure}

The second approach employs computational analysis, as illustrated in Fig.~\ref{GS2}. This method identifies the positions of sulci on the surface image of a brain and selects a random starting point. From this point, a circle is expanded until it encounters the closest sulcus. Multiple iterations yield an average radius, which is then quadrupled to estimate the gyral size. Notably, this approach generally produces slightly smaller gyral sizes compared to the characterization by width.

\subsubsection{Data Sources and Limitations}
Our analyses leverage brain images sourced from academic publications and the open-access website www.brainmuseum.org. The calculated gyral sizes across species are visually represented in Fig.~\ref{colored} and enumerated in Tables \ref{GS}, \ref{GS_2}, and \ref{GS_3}. Note that our data are sample-based. Specifically, when more than one sample is available for the same species, each sample is measured and presented independently. The limitations imposed by the number of available samples prevent us from providing an accurate analysis for each individual species. Instead, we focus on discerning patterns between gyral size and other factors across multiple species. Additionally, factors such as image quality and scale bar size could affect the accuracy of measurements. We predominantly employed computational methods (as shown in Fig.~\ref{GS2}), reverting to width characterization (as illustrated in Fig.~\ref{GS1}) when limitations necessitated it.

\subsection{Lissencephalic Brains}
\begin{figure}[htbp]
\centering
\includegraphics[width=5.6cm]{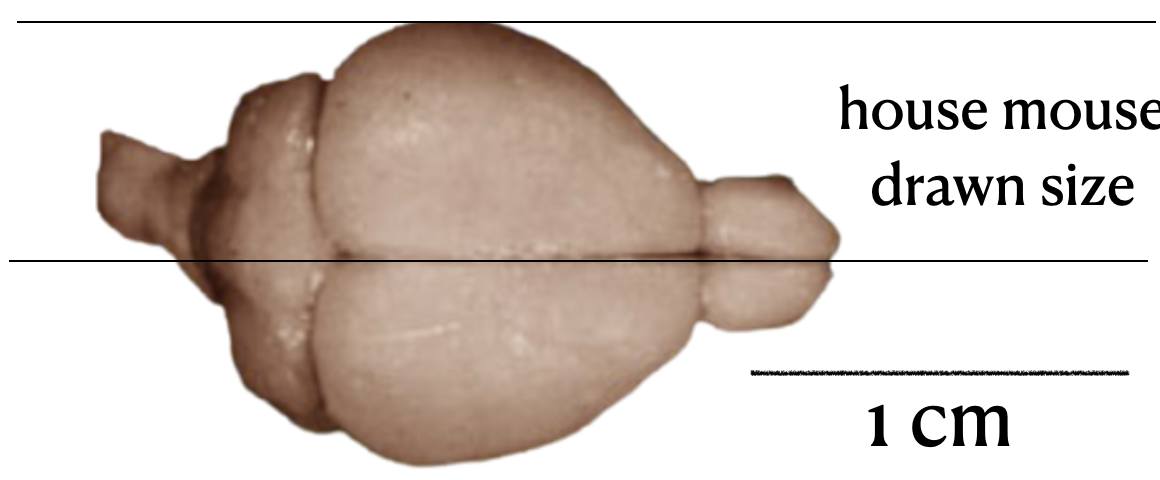}
\caption{
Measuring the hemisphere size of a house mouse's brain \cite{www.brainmuseum.org}. This measurement approach is applied to other lissencephalic brains. The data are integrated into Fig.~\ref{colored} and Table~\ref{smooth}.
}
\label{Lis}
\end{figure}

Lissencephalic brains, which are devoid of gyri and feature smooth surfaces, serve as comparators for gyrencephalic brains. We evaluate them using hemisphere size as a surrogate measure for gyral size in gyrencephalic brains. An example is provided in Fig.~\ref{Lis}. The measurements are listed in Table~\ref{smooth} and are also plotted in Fig.~\ref{colored} for comparison purposes. Significantly, lissencephalic features are predominantly associated with smaller brains. For larger brains, they begin to exhibit discernible sulci, as evident in the eastern gray squirrel \cite{www.brainmuseum.org}, European (domestic) rabbit \cite{www.brainmuseum.org}, and woodchuck (groundhog) \cite{www.brainmuseum.org}.

\subsection{Other Brain Types}
\begin{figure}[htbp]
\centering
\includegraphics[width=5.6cm]{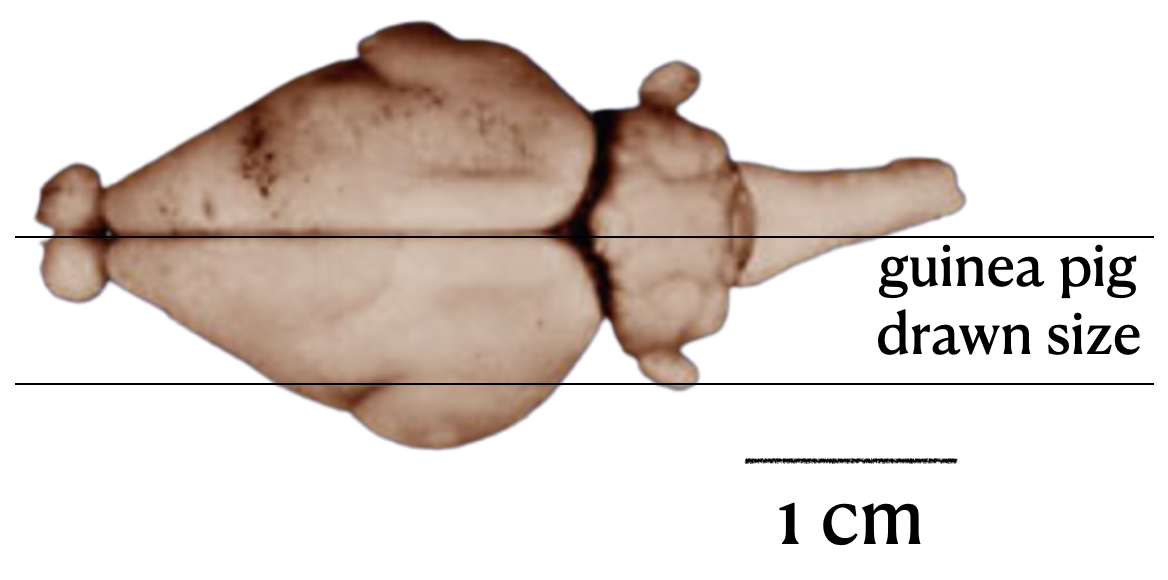}
\includegraphics[width=6.6cm]{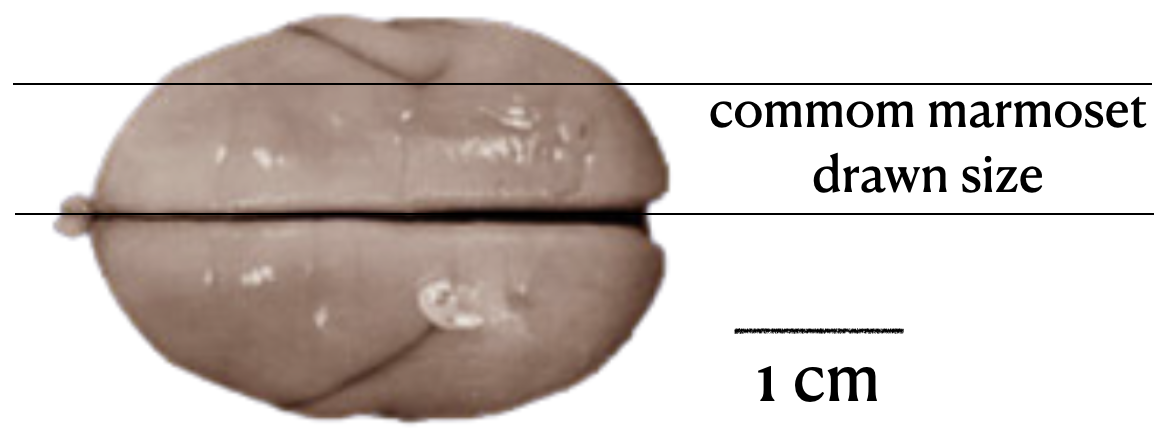}
\caption{
Quasi-gyrencephalic brains of a guinea pig \cite{www.brainmuseum.org} and a common marmoset \cite{www.brainmuseum.org}. These brains each feature a single sulcus in a hemisphere. The base of the sulcus serves as a reference point for our measurements, which contribute to the data presented in Fig.~\ref{colored}. These measurements allow for comparisons with gyral sizes in gyrencephalic brains. This measurement approach is applied to other quasi-gyrencephalic brains as well. The measurements are visually represented in Fig.~\ref{colored} and tabulated in Table~\ref{tablehint}.
}
\label{quasi}
\end{figure}

\begin{figure}[htbp]
\centering
\includegraphics[width=5.6cm]{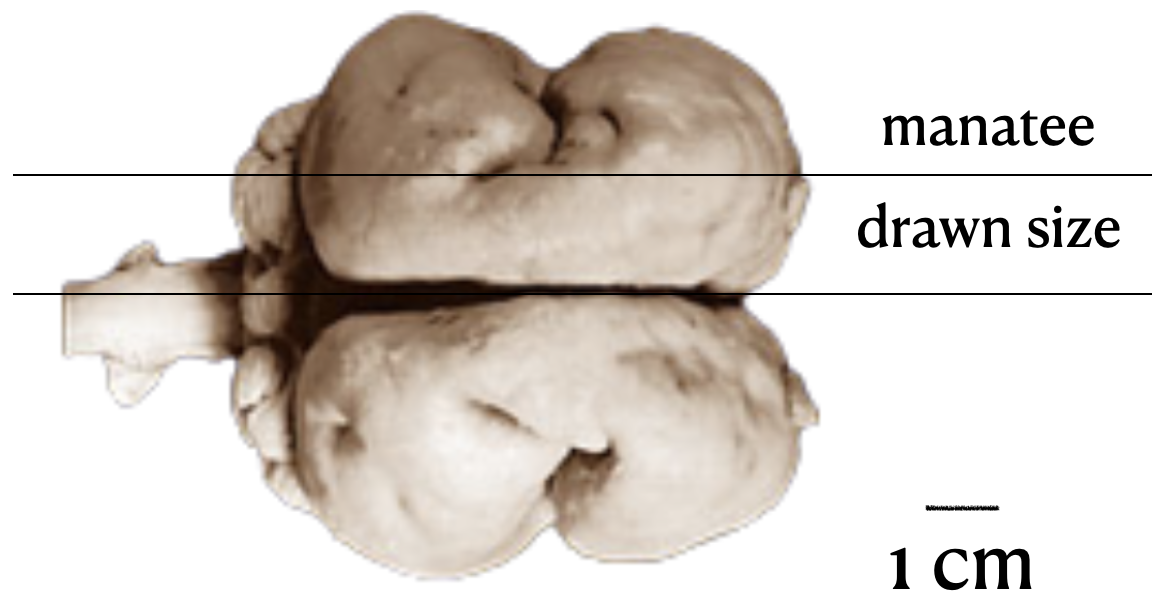}
\caption{
A manatee brain \cite{www.brainmuseum.org}, characterized by large subdivisions.  To obtain measurements that are comparable to gyral sizes in gyrencephalic brains, we apply the same measurement method used for quasi-gyrencephalic brains. These measurements are integrated into Fig.~\ref{colored} and Table~\ref{tablehint} for comprehensive analysis.
}
\label{hint}
\end{figure}

\begin{figure}
\centering
\includegraphics[width=8.6cm]{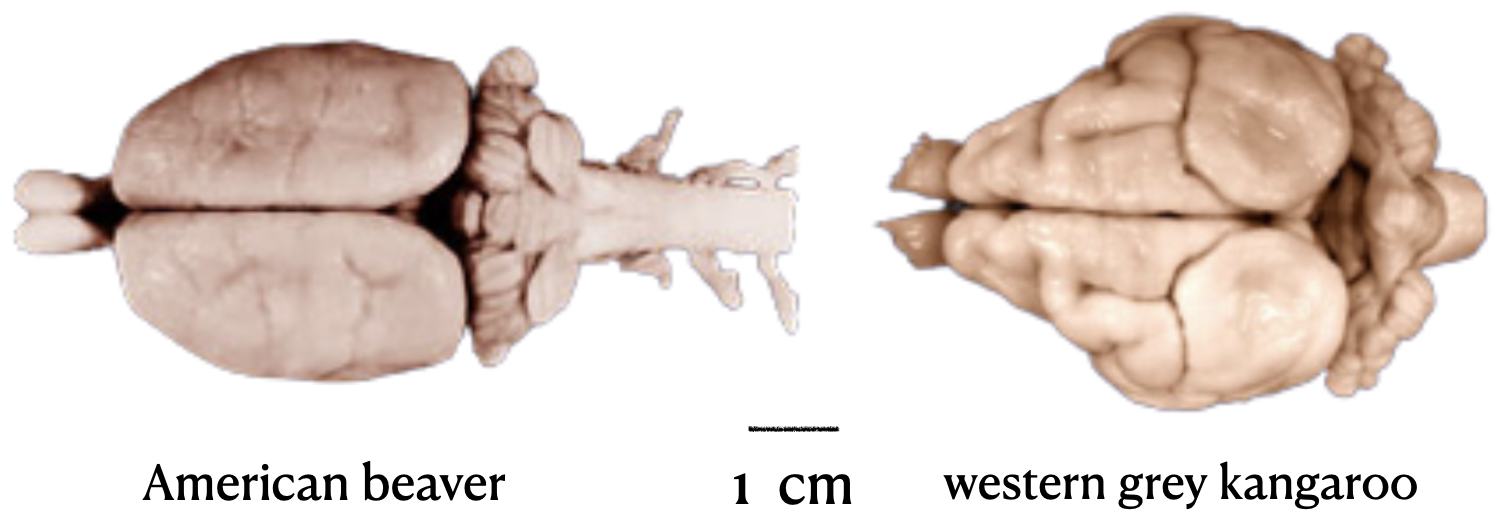}
\caption{
Challenging cases: the brains of the American beaver, which has cortical dimples, and the western grey kangaroo, which features irregular sulci \cite{www.brainmuseum.org}. These examples pose challenges for measuring equivalent gyral sizes. In the case of the American beaver, it remains ambiguous whether a cortical dimple can be classified as a sulcus. For the western grey kangaroo, the gyral sizes are not uniform, rendering the method outlined in Fig.\ref{GS1} unsuitable. Furthermore, the sparse number of sulci compromises the accuracy of the approach delineated in Fig.\ref{GS2}. Extending the methodology of Fig.\ref{GS2} to a 3-dimensional framework could offer a solution to these challenges. Additional examples of such challenging cases are compiled in Table~\ref{dimple} for brains with dimples and in Table~\ref{irregular} for those with irregular sulci.
}
\label{dimple_irregular}
\end{figure}

Some brains possess just a few, or even only one, sulcus in a hemisphere, as illustrated in Fig.~\ref{quasi}. We categorize these types of brains as quasi-gyrencephalic to distinguish them from classical lissencephalic and gyrencephalic varieties. For comparative analyses, we employ a specific metric, as depicted in Fig.~\ref{quasi}. This metric enables comparisons with both gyral sizes in gyrencephalic brains and hemisphere sizes in lissencephalic brains.
  The brains of manatees and dugongs exhibit unique structural characteristics, as demonstrated in Fig.~\ref{hint}. For these species, we make similar measurements to facilitate comparative analyses.
More complex cases are illustrated in Fig.~\ref{dimple_irregular}. For example, the American beaver's cortex features shallow dimples \cite{www.brainmuseum.org, 10.1159/000123818}. To manage this intricacy, one could extend the 2D method, as illustrated in Fig.~\ref{GS2}, to a 3D measurement framework if a 3D brain image is available. The brain of the western grey kangaroo presents comparable challenges due to its sparse and irregular sulci, rendering the methods demonstrated in Fig.~\ref{GS1} and Fig.~\ref{GS2} ineffective. 

\section{Results}
Our analysis reveals the relationship between gyral size and overall brain size, as well as in the categorization of different brain types. These findings not only corroborate previous research but also offer new insights into possible lifestyle influences on gyral dimensions.

\subsection{Gyral Size vs Brain Size}
Despite limitations in precision, our data about gyrencephalic brains, presented in Fig.~\ref{colored}, align with the findings of Heuer et al. \cite{10.1016/j.cortex.2019.04.011}. Specifically, we found little correlation between gyral sizes and overall brain sizes, particularly in larger mammals. This suggests a proportional relationship between cortical surface area and brain volume, supported by a power relation close to 1. Previous studies have reported similar power relations: 0.93 for mammals \cite{10.1126/science.166.3901.111}, 0.91 \cite{10.1159/000121313} or 0.89  \cite{10.1159/000118718} for terrestrial mammals, and 0.94 \cite{10.1159/000454797} for Delphinidae. These figures are closer to 1 than to the 2/3 expected if gyral and brain sizes were proportional. 

\subsection{Influence of Lifestyle on Gyral Sizes}
Our data in Fig.~\ref{colored} also suggest that gyral sizes vary widely, from 0.4 cm to 1.5 cm, potentially due to lifestyle factors. Species with narrower gyral sizes, such as ermines, goats, and bottlenose dolphins \cite{10.1038/s41598-021-00867-6}, often exhibit high-risk behaviors like aggressive hunting or head-ramming. In contrast, species like bowhead whale \cite{10.1002/ar.23991}, which have larger gyral size, generally live in environments with fewer head impact risks. Intriguingly, bowhead whales often use sea ice as a refuge from their main predators, killer whales \cite{10.1073/pnas.1911761117}.

\subsection{Classification of Brain Types Across Mammals}
Extending our analysis beyond gyrencephalic brains, Fig.~\ref{colored} indicates that we can categorize mammalian brains into three primary classifications: lissencephalic, quasi-gyrencephalic, and gyrencephalic. However, this classification scheme does not accommodate the challenging cases illustrated in Fig.~\ref{dimple_irregular}. For example, the American beaver and the western grey kangaroo serve as illustrative cases that blur the lines between these three categories. It may be necessary to introduce two additional categories in between, represented by the samples listed in Table~\ref{dimple} and Table~\ref{irregular}, respectively.

Notably, we observed a classification discrepancy concerning the brain of the nine-banded armadillo. One source, which features a pictorial atlas of coronal sections, shows that it is gyrencephalic, based on a brain size of 2.9 cm \cite{www.brainmuseum.org}. In contrast, another study classifies the armadillo's brain as lissencephalic, based on a brain size of 2.4 cm \cite{10.1111/ahe.12501}. This discrepancy may be attributed to variations in sample sizes. Indeed, our measurements suggest that both brain size and lifestyle factors influence brain gyrification patterns.

\section{Discussion}
\subsection{Risk of Head Impact}
To assess the risk of head impact, we consider two main factors: intensity and frequency. For example, beavers engage in controlled wood-cutting, which implies controlled intensity; this correlates with their near-lissencephalic brains. Killer whales, on the other hand, frequently use intense head-ramming during hunting, aligning with their narrow gyri \cite{10.1111/mms.12906}. Some species, like the capybara, face infrequent but potentially fatal head impacts when being hunted; this aligns with their gyrencephalic brains but not particularly narrow gyral sizes. Goats, known for head-ramming and susceptible to brain injury \cite{10.1007/s00401-022-02427-2}, have narrow gyri that fit their high-intensity lifestyle.

Atypical examples also offer insights. The Philippine flying lemurs  \cite{www.brainmuseum.org} are prone to accidental falls, aligning with their narrow gyral sizes. Beluga whales, despite not being known for high head impact risks, use intense echolocation signals \cite{10.1121/1.392341}. Their narrow gyral sizes may serve as a protective mechanism against potential acoustic harm.

\subsection{Intercepting Acoustic Waves}
Besides being actively generated by some animals for echolocation uses, acoustic waves can also result from head impacts due to the skull's rigidity.  These acoustic waves can be harmful to the brain \cite{10.3390/brainsci7060059, 10.1038/s41598-019-47295-1}. They induce oscillations in brain tissue \cite{10.1109/tuffc.2020.3022567}. Studies by Clayton et al. \cite{10.1098/rsif.2012.0325} and Okamoto et al. \cite{10.1177/1179069519840444} have demonstrated that internal membranes, such as the falx cerebri and the tentorium cerebelli, play crucial roles in reflecting and focusing shear waves within the brain. Simulations further underscore the significance of acoustic waves \cite{10.1007/s00707-009-0274-0, 10.1002/cnm.2881, 10.1007/978-3-319-28329-6_16}. The importance of acoustic waves in brain function is also evidenced by their therapeutic use in treating brain illnesses \cite{10.3389/fneur.2022.963849}.

\begin{figure}
\centering
\includegraphics[width=7.6cm]{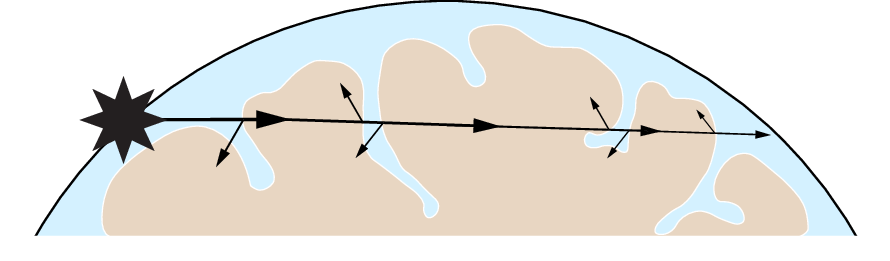}
\caption{
Sulci protecting the brain against acoustic waves: A head impact generates an acoustic wave pulse that propagates through the brain. The pulse encounters interfaces between brain tissue and CSF, leading to both reflection and refraction, thereby diminishing its energy and reducing its potential for harm.
}
\label{advantage}
\end{figure}

In Fig.~\ref{advantage}, an acoustic wave pulse is depicted as it propagates through a brain. Due to the distinct acoustic properties between cerebrospinal fluid (CSF) and brain tissue, the pulse experiences partial reflections at each boundary interface. On one hand, these reflections have the potential to amplify the damaging effects of the incoming wave, particularly in the vicinity of the boundary. On the other hand, these reflections also attenuate the wave's energy, rendering the transmitted wave less harmful. The overall impact of these reflections should be beneficial.

\begin{figure}
\centering
\includegraphics[width=7.6cm]{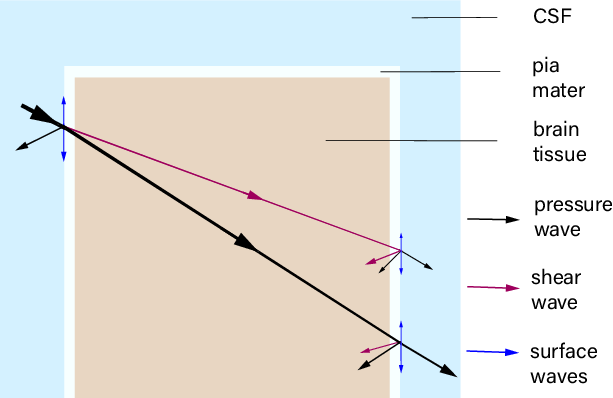}
\caption{
Schematic of an acoustic wave crossing a gyrus: The incoming pressure wave from the CSF encounters the interface between CSF and the pia mater, generating a variety of wave phenomena that further split upon reaching the gyrus's opposite boundary.
}
\label{pia}
\end{figure}

For simplicity, we focus on the interface between CSF and the pia mater, a thin membrane enveloping the brain tissue \cite{10.1111/j.1365-2990.1988.tb00862.x}. We omit additional boundaries and the distinction between gray and white matter. Some details of wave propagation can be summarized as follows [refer to Fig.~\ref{advantage} and Fig.~\ref{pia}]:

\begin{enumerate}
\item A head impact generates a pressure wave pulse in the CSF, which being a liquid, allows only the passage of pressure or P-waves.

\item Upon reaching a gyrus, the pulse spawns multiple phenomena: reflected and transmitted pressure waves, a transmitted shear or S-wave, and various surface waves \cite{10.1201/b12260}.

\item The transmitted pressure and shear waves traverse the gyrus, separating due to their different speeds \cite{10.1103/physrevapplied.8.044024}. At the opposite boundary, each wave splits again.

\item Both transmitted pressure waves then advance through the CSF to the next gyrus.
\end{enumerate}


The role of the skull in head impact scenarios is also noteworthy. Head impacts generate surface waves, such as Lamb waves, that propagate along the skull \cite{10.1039/d0lc00887g}. In skulls that are nearly spherical, like those of humans, these waves can focus at the antipode. This serves as a secondary source of acoustic waves and may contribute to contrecoup injuries \cite{10.3109/02688697.2010.548879}. This phenomenon bears a resemblance to the antipodal effect observed in geology \cite{10.1007/bf00577875,10.3389/feart.2022.827252}. On one hand, spherical skulls offer the advantage of providing more volume for a larger brain. On the other hand, less symmetric skulls may minimize the antipodal focusing effect, thereby reducing associated risks. These factors could have evolutionary implications on skull shape, as different species may prioritize different aspects depending on their lifestyle and environmental pressures.

\subsection{Gyral Size and Brain Size}
\begin{figure}[htbp]
\centering
\includegraphics[width=7.6cm]{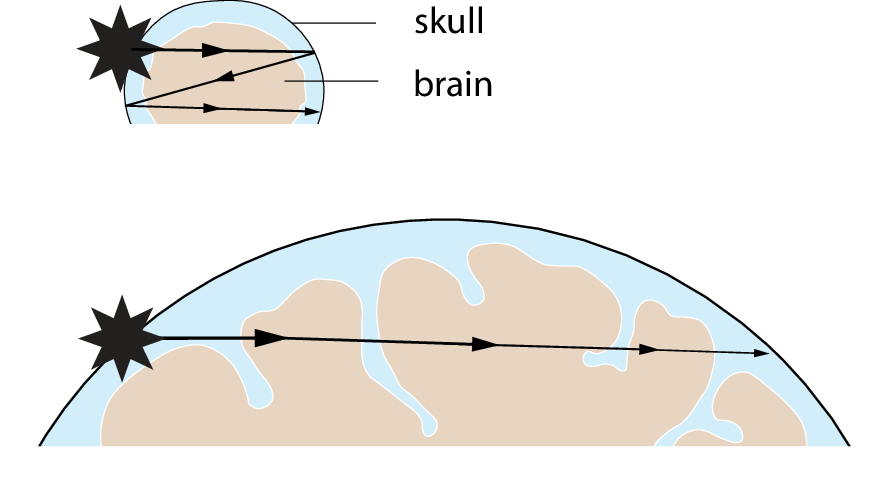}
\caption{
Comparison of a small lissencephalic brain and a large gyrencephalic brain. The entire hemisphere of the small brain is functionally analogous in size to a single gyrus in the larger brain. Both brains offer similar levels of protection against acoustic waves. An acoustic wave traversing the small brain is functionally equivalent to one moving through a series of gyri in the larger brain.
}
\label{smallbrain}
\end{figure}

As illustrated in Fig.~\ref{smallbrain}, both small and large brains can provide comparable levels of protection against acoustic waves. The figure contrasts a small, lissencephalic brain with a larger, gyrencephalic one, underscoring that the entire hemisphere of the small brain is functionally analogous in size to a single gyrus in the larger brain. Because of this similarity, both types of brains dissipate acoustic wave energy at comparable rates due to reflections at the CSF-brain tissue boundary and the skull. Consequently, they have an equal likelihood of either sustaining or avoiding damage from such waves.

This observation may illuminate why smaller brains are often lissencephalic. Should a small brain increase in size while maintaining the same level of protection, it would likely transition to a gyrencephalic structure with gyri of sizes similar to those in the larger brain. Theoretically, assuming all other factors remain constant, the same size of gyri would be necessary for identical levels of protection against acoustic waves, regardless of the brain's overall size. This suggests that the observed independence between gyral size and brain size, as highlighted in Fig.~\ref{colored}, could indicate that acoustic waves might be the primary risk factor for brain safety in mammals, or that the main mechanism through which head impacts cause brain damage is via acoustic waves.

Furthermore, this analysis offers insights into various patterns of gyrification. Complex gyrification is more likely observed in species with larger brains, or those facing a high risk of head impacts due to lifestyle factors or intense echolocation signals. Conversely, lissencephalic patterns are generally found in species with smaller brains or a lower risk of head impacts. Intermediate patterns exist between these two extremes.

\subsection{Mitigating Acceleration}
Brain sulci may also serve to mitigate the acceleration caused by head impact. The in vivo brain exhibits a slightly higher density compared to the cerebrospinal fluid (CSF) found within the sulci. Materials with differing densities respond differently when subjected to acceleration \cite{10.1016/s0046-8177(80)80136-5}. This differential response can have both beneficial and detrimental effects. Kornguth et al. \cite{10.3390/brainsci7120164} discussed the fluidity of CSF and introduced the concept of a "water hammer injury effect." According to this theory, a head impact propels CSF into the sulci, focusing the majority of the force of the CSF impulse at the base of the sulcus. Both observational \cite{10.1093/brain/aws307,10.1001/jama.2017.8334} and simulation studies \cite{10.1093/brain/aww317} support the notion that damage is more likely to occur in sulcal regions. The hammer effect is especially pronounced when a sulcus is oriented parallel to the direction of acceleration.

\begin{figure}[htbp]
\centering
\includegraphics[width=8.6cm]{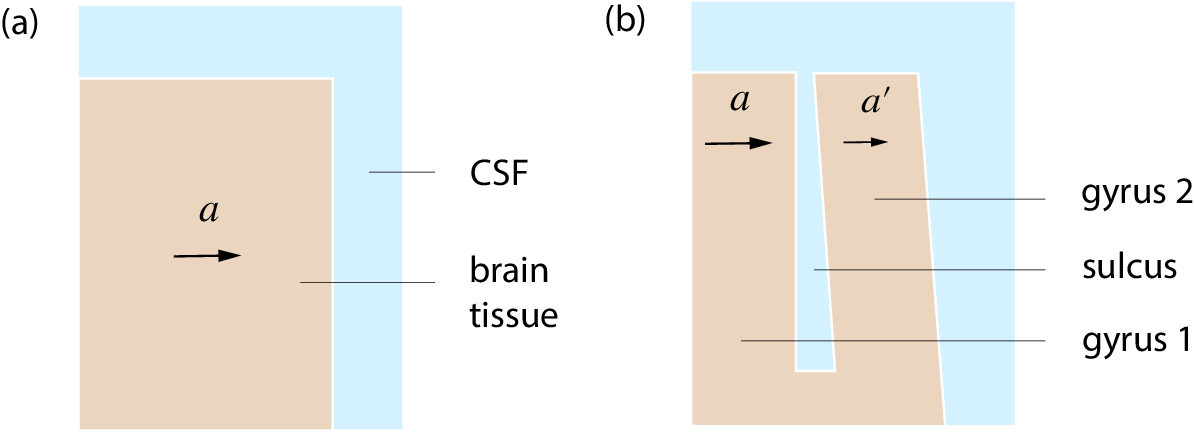}
\caption{
Sulcus protecting gyrus against acceleration: (a) This segment represents part of a single gyrus, bordered by CSF on both sides. Upon head impact, a force propagates through adjacent brain regions to this segment, resulting in an acceleration denoted by \(a\). (b) A sulcus is introduced, creating a second gyrus separated from the original gyrus by a CSF-filled sulcus. While the remaining part of the first gyrus experiences the same acceleration \(a\) as before, the second gyrus undergoes a reduced acceleration, \(a' < a\), due to the expulsion of CSF from the sulcus.
}
\label{acceleration}
\end{figure}

Conversely, when a sulcus is oriented perpendicularly to the direction of acceleration, the density difference between the brain tissue and CSF can act as a protective mechanism, as illustrated in Fig.~\ref{acceleration}. A head impact generates a force that propagates through the brain, causing adjacent brain regions to compress. This compression forces the less dense CSF out of the sulci, thereby reducing the compressive force and allowing the adjacent gyrus more time to respond, effectively mitigating the acceleration. However, if the acceleration is excessively strong, adjacent gyri may collide, resulting in cortical surface damage. This mechanism is analogous to how CSF cushions the impact between the brain and the skull \cite{10.1080/02699052.2018.1502470}.

\subsection{Species-Specific Gyrification Patterns}
Both Fig.~\ref{advantage} and Fig.~\ref{acceleration} demonstrate that sulci oriented perpendicular to the direction of an incoming impact are most effective at mitigating it. This observation may provide insights into gyrification patterns. For example, species like cats, depicted in Fig.~\ref{GS1}, and the Ganges River dolphin \cite{10.1159/000454797}, possess parallel sulci located in the central regions of their brains. This specific orientation could be related to particular anatomical traits. Cats have forward-facing eyes, while the Ganges River dolphin has an elongated snout. As a result, both species can more readily avoid head impacts from the forward direction. This may explain why their sulci in the central regions are oriented to mitigate impacts primarily coming from the sides. In contrast, for goats, which have integrated horn structures \cite{10.1111/j.1469-7998.1995.tb05137.x}, a head impact coming from the horn can be dispersed into their entire skulls. This aligns with their brain gyrification patterns, which do not seem to prioritize any specific directions.

\subsection{Evolutionary Trade-offs in Brain Gyrification}
Manatees and dugongs, renowned for their unique large brain subdivisions, differ significantly from their close relatives, the elephants, which have gyrencephalic brains \cite{10.1002/cne.23046}. These species are believed to have diverged from a common gyrencephalic ancestor, especially as manatees and dugongs adapted to a peaceful aquatic lifestyle \cite{10.1017/cbo9781139013277}. Over time, manatees evolved to have fewer sulci but increased their cortical thickness to around 4 mm \cite{10.1159/000115866}. This exceeds the human average of 3.4 mm \cite{10.1038/nrn3707} and is considerably thicker than most other animals, which often have a cortical thickness under 2 mm. This increased thickness is likely an evolutionary adaptation to maintain or even augment the neuron count, offsetting the loss in cortical surface area. Although greater surface area, often due to gyrification, is traditionally associated with higher neuron counts, manatees may achieve comparable neuron count through increased cortical thickness.

\begin{figure}
\centering
\includegraphics[width=7.6cm]{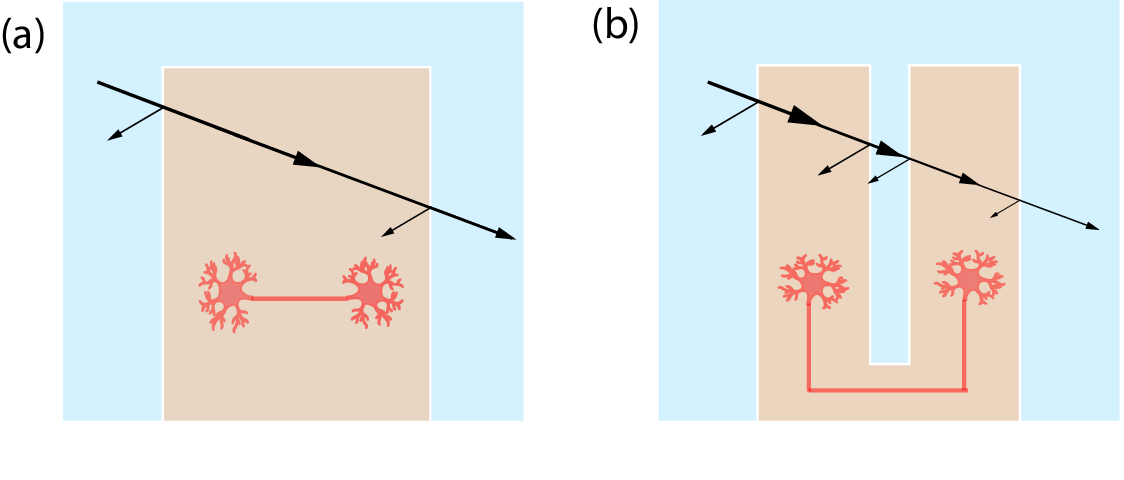}
\caption{
Schematic illustrating the trade-offs tied to having a sulcus. (a) Without sulci, neural pathways are direct but offer less acoustic attenuation. (b) With a sulcus, neural pathways are circuitous, but acoustic waves are better attenuated. The need for effective neural connections is universal, whereas the need for protective adaptations varies by species, especially their lifestyle associated with risk factors. Thus, the presence of sulci, or gyral size, depends mainly on lifestyles.
}
\label{sulcus}
\end{figure}

\begin{figure*}[htbp]
\centering
\resizebox{\textwidth}{!}{
\includegraphics[width=18.6cm]{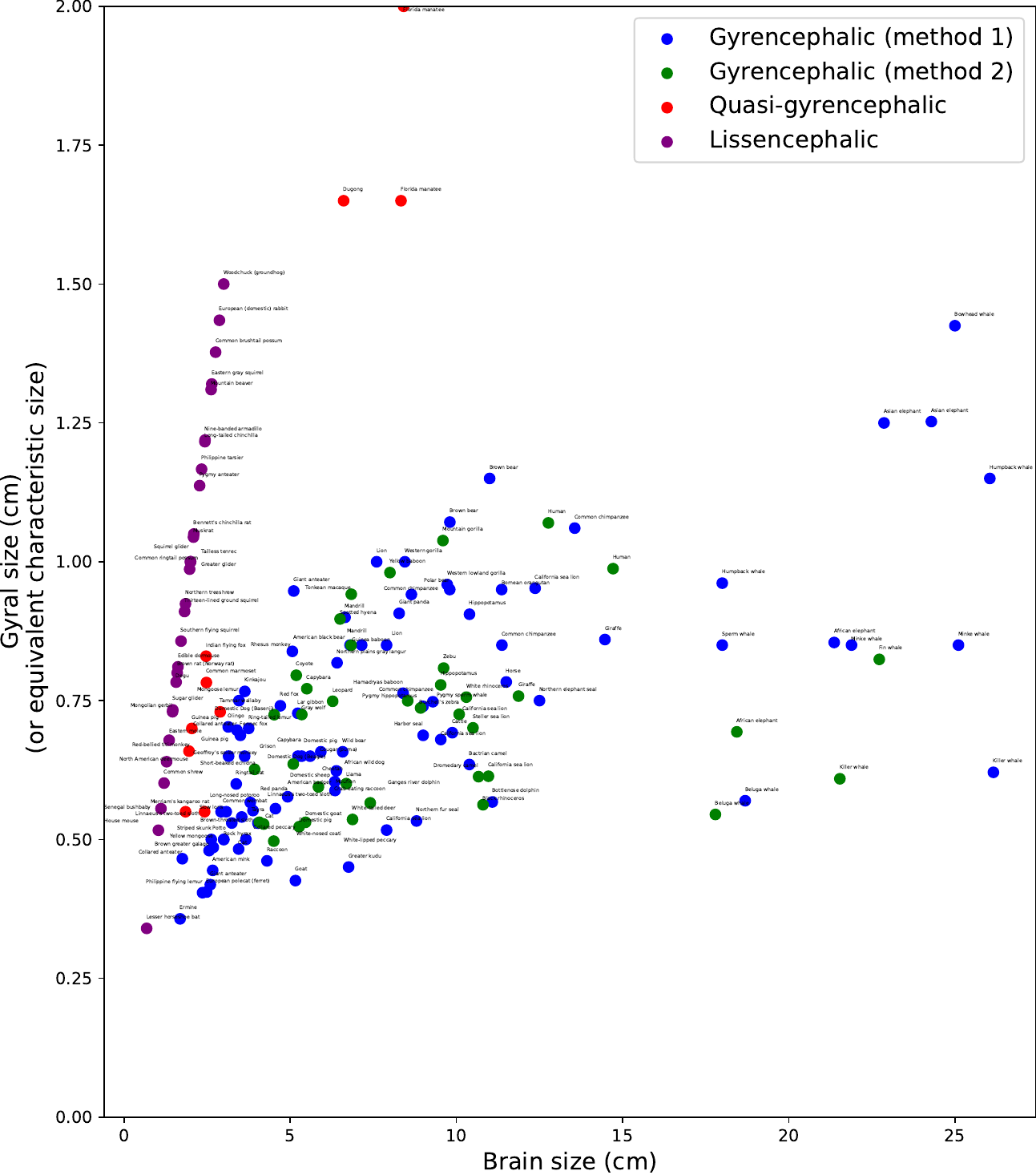}
}
\caption{
Plot of gyral size versus brain size for gyrencephalic brains. Blue and green points indicate gyral sizes measured using different methods, as illustrated in Fig.~\ref{GS1} and Fig.~\ref{GS2}, respectively. For comparison, two other types of brains are also included. Purple points represent lissencephalic brains, with hemisphere sizes measured as depicted in Fig.~\ref{Lis}. Red points denote quasi-gyrencephalic brains, measured as shown in Fig.~\ref{quasi} and Fig.~\ref{hint}. Data are collected from academic publications and www.brainmuseum.org. For species with multiple images, each image is measured individually. When possible, brain sizes are measured along the lateral axis, encompassing both hemispheres.
}
\label{colored}
\end{figure*}

The reduction in the number of sulci in the brains of manatees suggests evolutionary trade-offs in brain gyrification. One possible trade-off could be that sulci may compromise efficient neural connectivity, as illustrated in Fig.~\ref{sulcus}. For species with low risks of head impact—such as manatees—a lissencephalic structure, devoid of sulci, might facilitate more direct and efficient neural pathways. Conversely, species like humans, who must balance the risks of head impact with the need for neural efficiency, may benefit from moderate gyrification.

\section{Conclusions}
In summary, we demonstrate that gyral size does not consistently correlate with overall brain size, especially in species with larger brains. Our study also introduces two key mechanisms by which gyrification may confer protective benefits: cushioning the brain through fluid-filled sulci and attenuating the impact of harmful acoustic waves. However, our findings also reveal a complex trade-off: while sulci may offer some level of protection, they could simultaneously impede efficient neural connections. This highlights the intricate balance between the universal need for effective neural connections and the species-specific requirements for protective adaptations, which vary based on lifestyle and associated risks. Our study has limitations, including the need for a larger sample size and higher-resolution brain images. Future research may focus on further validating gyral size as a useful parameter and collecting more comprehensive data on gyral sizes. Overall, our findings open new avenues for understanding the biomechanical and evolutionary pressures that influence brain morphology, without ruling out the possibility that gyri may serve other functions.

\begin{table*}[htbp]
\caption{
Estimation of gyral sizes across species using method 1 and method 2, as described in Fig.~\ref{GS1} and Fig.~\ref{GS2}, respectively. Data are sourced from academic publications and www.brainmuseum.org. When several brain images are available for the same species, each image is measured individually. Most measurements of brain sizes are taken along the lateral axis.
\label{GS}}
\centering
\resizebox{\textwidth}{!}{
\begin{tabular}{c|lllcc}
\hline
\textbf{Gyral size (cm)} & \textbf{Common name} & \textbf{Species name} & \textbf{Order} & \textbf{Brain size (cm)} & \textbf{Method} \\
\hline
\hline
0.3--0.4 & Ermine\cite[\#63-10]{www.brainmuseum.org}& {\it Mustela erminea }& Carnivora & 1.7& 1 \\
\hline
0.4--0.5 & Collared anteater\cite[Fig.~6]{10.1111/ahe.12501}& {\it Tamandua tetradactyla }& Pilosa & 1.8& 1 \\
 & Philippine flying lemur\cite[\#63-271]{www.brainmuseum.org}& {\it Cynocephalus volans }& Dermoptera & 2.4& 1 \\
 & European polecat (ferret)\cite[\#58-323]{www.brainmuseum.org}& {\it Mustela putorius }& Carnivora & 2.5& 1 \\
 & Brown greater galago\cite[\#62-172]{www.brainmuseum.org}& {\it Otolemur crassicaudatus }& Primates & 2.6& 1 \\
 & Giant anteater\cite[Fig.~6]{10.1111/ahe.12501}& {\it Myrmecophaga tridactyla }& Pilosa & 2.6& 1 \\
 & Striped skunk\cite[\#63-113]{www.brainmuseum.org}& {\it Mephitis mephitis }& Carnivora & 2.6& 1 \\
 & American mink\cite[\#58-324]{www.brainmuseum.org}& {\it Neovison vison }& Carnivora & 2.7& 1 \\
 & Yellow mongoose\cite[\#61-766]{www.brainmuseum.org}& {\it Cynictis penicillata }& Carnivora & 2.7& 1 \\
 & Rock hyrax\cite[\#62-635]{www.brainmuseum.org}& {\it Procavia capensis }& Hyracoidea & 3.0& 1 \\
 & Potto\cite[\#62-441]{www.brainmuseum.org}& {\it Perodicticus potto }& Primates & 3.0& 1 \\
 & Cat\cite[Fig.~7]{10.3389/fnana.2011.00029}& {\it Felis catus }& Carnivora & 3.4& 1 \\
 & White-nosed coati\cite[\#58-360]{www.brainmuseum.org}& {\it Nasua narica }& Carnivora & 3.7& 1 \\
 & Raccoon\cite[\#57-88]{www.brainmuseum.org}& {\it Procyon lotor }& Carnivora & 4.3& 1 \\
 & White-lipped peccary\cite[Fig.~7]{10.3389/fnana.2011.00029}& {\it Tayassu pecari  }& Artiodactyla & 4.5& 2 \\
 & Goat\cite[Fig.~7]{10.3389/fnana.2011.00029}& {\it Capra aegagrus hircus }& Artiodactyla & 5.2& 1 \\
 & Greater kudu\cite[Fig.~7]{10.3389/fnana.2011.00029}& {\it Tragelaphus strepsiceros }& Artiodactyla & 6.8& 1 \\
\hline
0.5--0.6 & Slow loris\cite[\#62-181]{www.brainmuseum.org}& {\it Nycticebus coucang }& Primates & 2.9& 1 \\
 & Common wombat\cite[\#64-11]{www.brainmuseum.org}& {\it Vombatus ursinus }& Diprotodontia & 3.1& 1 \\
 & Brown-throated sloth\cite[Fig.~4]{10.1111/ahe.12501}& {\it Bradypus variegatus }& Pilosa & 3.2& 1 \\
 & Ringtail cat\cite[\#62-106]{www.brainmuseum.org}& {\it Bassariscus astutus }& Carnivora & 3.4& 1 \\
 & Linnaeus's two-toed sloth\cite[\#61-98]{www.brainmuseum.org}& {\it Choloepus didactylus }& Pilosa & 3.5& 1 \\
 & Red panda\cite[\#60-121]{www.brainmuseum.org}& {\it Ailurus fulgens }& Carnivora & 3.8& 1 \\
 & Linnaeus's two-toed sloth\cite[Fig.~1]{10.1111/ahe.12501}& {\it Choloepus didactylus }& Pilosa & 3.9& 1 \\
 & Cat\cite[\#60-330]{www.brainmuseum.org}& {\it Felis catus }& Carnivora & 4.0& 1 \\
 & Tayra\cite[\#69-56]{www.brainmuseum.org}& {\it Eira barbara }& Carnivora & 4.1& 2 \\
 & Collared peccary\cite[\#63-445]{www.brainmuseum.org}& {\it Pecari tajacu }& Artiodactyla & 4.2& 2 \\
 & Crab-eating raccoon\cite[\#68-312]{www.brainmuseum.org}& {\it Procyon cancrivorus }& Carnivora & 4.6& 1 \\
 & American badger\cite[\#63-127]{www.brainmuseum.org}& {\it Taxidea taxus }& Carnivora & 4.9& 1 \\
 & Domestic pig\cite[\#61-361]{www.brainmuseum.org}& {\it Sus scrofa domesticus }& Artiodactyla & 5.3& 2 \\
 & Domestic goat\cite[\#59-55]{www.brainmuseum.org}& {\it Capra hircus domestica }& Artiodactyla & 5.5& 2 \\
 & Domestic sheep\cite[\#61-693]{www.brainmuseum.org}& {\it Ovis aries }& Artiodactyla & 5.8& 2 \\
 & Mouflon\cite[Fig.~7]{10.3389/fnana.2011.00029}& {\it Ovis orientalis }& Artiodactyla & 6.3& 1 \\
 & White-tailed deer\cite[\#67-81]{www.brainmuseum.org}& {\it Odocoileus virginianus }& Artiodactyla & 6.9& 2 \\
 & Ganges river dolphin\cite[Fig.~19]{10.1159/000454797}& {\it Platanista gangetica }& Cetacea & 7.4& 2 \\
 & California sea lion\cite[Fig.~18]{10.1017/S1464793106007019}& {\it Zalophus californianus }& Carnivora & 7.9& 1 \\
 & Northern fur seal\cite[\#61-511]{www.brainmuseum.org}& {\it Callorhinus ursinus }& Carnivora & 8.8& 1 \\
 & Black rhinoceros \cite[Fig.~1]{10.3389/fnana.2017.00074}& {\it Diceros bicornis  }& Perissodactyla & 11& 2 \\
 & Bottlenose dolphin\cite[Fig.~2]{10.1159/000063731}& {\it Tursiops truncatus }& Cetacea & 11& 1 \\
 & Beluga whale\cite[Fig.~19]{10.1159/000454797}& {\it Delphinapterus leucas }& Cetacea & 18& 2 \\
 & Beluga whale\cite[Fig.~1]{10.1002/ar.10183}& {\it Delphinapterus leucas }& Cetacea & 19& 1 \\
\hline\end{tabular}
}
\end{table*}

\begin{table*}[htbp]
\caption{Continuation of Table~\ref{GS}. \label{GS_2}}
\centering
\resizebox{\textwidth}{!}{
\begin{tabular}{c|lllcc}
\hline
\textbf{Gyral size (cm)} & \textbf{Common name} & \textbf{Species name} & \textbf{Order} & \textbf{Brain size (cm)} & \textbf{Method} \\
\hline
\hline
0.6--0.7 & Red-bellied titi monkey\cite[\#70-355]{www.brainmuseum.org}& {\it Callicebus moloch }& Primates & 3.1& 1 \\
 & Collared anteater\cite[\#61-93]{www.brainmuseum.org}& {\it Tamandua tetradactyla }& Pilosa & 3.4& 1 \\
 & Fennec fox\cite[\#63-388]{www.brainmuseum.org}& {\it Vulpes zerda }& Carnivora & 3.5& 1 \\
 & Grison\cite[\#69-11]{www.brainmuseum.org}& {\it Galictis vittata }& Carnivora & 3.6& 1 \\
 & Ring-tailed lemur\cite[\#62-442]{www.brainmuseum.org}& {\it Lemur catta }& Primates & 3.8& 1 \\
 & Short-beaked echidna\cite[\#64-232]{www.brainmuseum.org}& {\it Tachyglossus aculeatus }& Monotremata & 3.9& 2 \\
 & Domestic Dog (Beagle)\cite[\#59-326]{www.brainmuseum.org}& {\it Canis lupus familiaris }& Carnivora & 5.1& 2 \\
 & Capybara\cite[Fig.~2]{10.1159/000123814}& {\it Hydrochoerus hydrochaeris }& Rodentia & 5.2& 1 \\
 & Geoffroy's spider monkey\cite[\#63-18]{www.brainmuseum.org}& {\it Ateles geoffroyi }& Primates & 5.3& 1 \\
 & Cougar (puma)\cite[\#60-206]{www.brainmuseum.org}& {\it Puma concolor }& Carnivora & 5.6& 1 \\
 & Domestic pig\cite[Fig.~7]{10.1002/cne.25461}& {\it Sus scrofa domesticus  }& Artiodactyla & 5.9& 1 \\
 & Cheetah\cite[Fig.~7]{10.3389/fnana.2011.00029}& {\it Acinonyx jubatus }& Carnivora & 6.3& 1 \\
 & African wild dog\cite[Fig.~3]{10.1002/cne.24999}& {\it Lycaon pictus }& Carnivora & 6.4& 1 \\
 & Wild boar\cite[Fig.~7]{10.1002/cne.25461}& {\it Sus scrofa }& Artiodactyla & 6.6& 1 \\
 & Llama\cite[\#65-139]{www.brainmuseum.org}& {\it Lama glama }& Artiodactyla & 6.7& 2 \\
 & Harbor seal\cite[\#61-515]{www.brainmuseum.org}& {\it Phoca vitulina }& Carnivora & 9.0& 1 \\
 & California sea lion\cite[Fig.~1]{10.1002/ar.20937}& {\it Zalophus californianus }& Carnivora & 9.5& 1 \\
 & Cattle\cite[Fig.~1]{10.1371/journal.pone.0154580}& {\it Bos taurus }& Artiodactyla & 9.9& 1 \\
 & Bactrian camel\cite[Fig.~2]{Camelus}& {\it Camelus bactrianus }& Artiodactyla & 10& 1 \\
 & Dromedary camel\cite[\#60-227]{www.brainmuseum.org}& {\it Camelus dromedarius }& Artiodactyla & 11& 2 \\
 & California sea lion\cite[Fig.~7]{10.1002/cne.23984}& {\it Zalophus californianus }& Carnivora & 11& 2 \\
 & African elephant\cite[\#85-33]{www.brainmuseum.org}& {\it Loxodonta africana }& Proboscidea & 18& 2 \\
 & Killer whale\cite[Fig.~1]{10.1017/S1464793106007019}& {\it Orcinus orca }& Cetacea & 22& 2 \\
 & Killer whale\cite[Fig.~19]{10.1159/000454797}& {\it Orcinus orca }& Cetacea & 26& 1 \\
\hline
0.7--0.8 & Olingo\cite[\#62-113]{www.brainmuseum.org}& {\it Bassaricyon gabbii }& Carnivora & 3.1& 1 \\
 & Mongoose lemur\cite[\#64-69]{www.brainmuseum.org}& {\it Eulemur mongoz }& Primates & 3.5& 1 \\
 & Kinkajou\cite[\#58-365]{www.brainmuseum.org}& {\it Potos flavus }& Carnivora & 3.6& 1 \\
 & Domestic Dog (Basenji)\cite[\#66-165]{www.brainmuseum.org}& {\it Canis lupus familiaris }& Carnivora & 4.5& 2 \\
 & Red fox\cite[\#63-392]{www.brainmuseum.org}& {\it Vulpes vulpes }& Carnivora & 4.7& 1 \\
 & Coyote\cite[\#62-301]{www.brainmuseum.org}& {\it Canis latrans }& Carnivora & 5.2& 2 \\
 & Lar gibbon\cite[\#60-142]{www.brainmuseum.org}& {\it Hylobates lar }& Primates & 5.2& 1 \\
 & Gray wolf\cite[\#71-26]{www.brainmuseum.org}& {\it Canis lupus }& Carnivora & 5.3& 2 \\
 & Capybara\cite[\#62-621]{www.brainmuseum.org}& {\it Hydrochoerus hydrochaeris }& Rodentia & 5.5& 2 \\
 & Leopard\cite[\#63-261]{www.brainmuseum.org}& {\it Panthera pardus }& Carnivora & 6.3& 2 \\
 & Hamadryas baboon\cite[Fig.~1]{10.1002/cne.903110109}& {\it Papio hamadryas }& Primates & 8.4& 1 \\
 & Common chimpanzee\cite[Fig.~7]{10.3389/fnana.2011.00029}& {\it Pan troglodytes }& Primates & 8.5& 2 \\
 & Burchell's zebra\cite[\#61-820]{www.brainmuseum.org}& {\it Equus quagga burchellii }& Perissodactyla & 8.9& 2 \\
 & Pygmy hippopotamus\cite[Fig.~1]{10.1002/ar.22875}& {\it Hexaprotodon liberiensis }& Artiodactyla & 9.0& 1 \\
 & Pygmy sperm whale\cite[Fig.~2]{10.1002/ar.23991}& {\it Kogia breviceps }& Cetacea & 9.3& 1 \\
 & Hippopotamus\cite[Fig.~2]{10.1002/cne.23930}& {\it Hippopotamus amphibius }& Artiodactyla & 9.5& 2 \\
 & California sea lion\cite[\#62-294]{www.brainmuseum.org}& {\it Zalophus californianus }& Carnivora & 10& 2 \\
 & White rhinoceros\cite[Fig.~2]{10.3389/fnana.2017.00074}& {\it Ceratotherium simum }& Perissodactyla & 10& 2 \\
 & Steller sea lion\cite[\#61-513]{www.brainmuseum.org}& {\it Eumetopias jubatus }& Carnivora & 10& 2 \\
 & Horse\cite[Fig.~1]{10.1159/000356527}& {\it Equus caballus }& Perissodactyla & 12& 1 \\
 & Giraffe\cite[Fig.~7]{10.3389/fnana.2011.00029}& {\it Giraffa camelopardalis }& Artiodactyla & 12& 2 \\
 & Northern elephant seal\cite[Fig.~9]{10.1002/cne.24188}& {\it Mirounga angustirostris }& Carnivora & 12& 1 \\
\hline\end{tabular}
}
\end{table*}

\begin{table*}[htbp]
\caption{Final part of Table~\ref{GS}. \label{GS_3}}
\centering
\resizebox{\textwidth}{!}{
\begin{tabular}{c|lllcc}
\hline
\textbf{Gyral size (cm)} & \textbf{Common name} & \textbf{Species name} & \textbf{Order} & \textbf{Brain size (cm)} & \textbf{Method} \\
\hline
\hline
0.8--0.9 & Rhesus monkey\cite[\#69-307]{www.brainmuseum.org}& {\it Macaca mulatta }& Primates & 5.1& 1 \\
 & Northern plains gray langur\cite[\#62-179]{www.brainmuseum.org}& {\it Semnopithecus entellus }& Primates & 6.4& 1 \\
 & Spotted hyena\cite[\#64-352]{www.brainmuseum.org}& {\it Crocuta crocuta }& Carnivora & 6.5& 2 \\
 & Mandrill\cite[\#63-310]{www.brainmuseum.org}& {\it Mandrillus sphinx }& Primates & 6.7& 1 \\
 & Mandrill\cite[Fig.~7]{10.3389/fnana.2011.00029}& {\it Mandrillus sphinx  }& Primates & 6.8& 1 \\
 & American black bear\cite[\#71-256]{www.brainmuseum.org}& {\it Ursus americanus }& Carnivora & 6.8& 2 \\
 & Guinea baboon\cite[\#64-12]{www.brainmuseum.org}& {\it Papio papio }& Primates & 7.2& 1 \\
 & Lion\cite[Fig.~7]{10.3389/fnana.2011.00029}& {\it Panthera leo }& Carnivora & 7.9& 1 \\
 & Zebu\cite[\#64-322]{www.brainmuseum.org}& {\it Bos taurus indicus }& Artiodactyla & 9.6& 2 \\
 & Common chimpanzee\cite[Fig.~1]{10.1002/ar.a.20256}& {\it Pan troglodytes }& Primates & 11& 1 \\
 & Giraffe\cite[Fig.~1]{10.1002/ar.23593}& {\it Giraffa camelopardalis }& Artiodactyla & 14& 1 \\
 & Sperm whale\cite[Fig.~2]{10.1002/ar.23991}& {\it Physeter macrocephalus }& Cetacea & 18& 1 \\
 & African elephant\cite[Fig.~1]{10.1002/cne.23817}& {\it Loxodonta africana }& Proboscidea & 21& 1 \\
 & Minke whale\cite[Fig.~19]{10.1159/000454797}& {\it Balaenoptera acutorostrata }& Cetacea & 22& 1 \\
 & Fin whale\cite[Fig.~1]{10.1002/ar.20407}& {\it Balaenoptera physalus }& Cetacea & 23& 2 \\
 & Minke whale\cite[Fig.~4]{10.1016/S0165-02700200182-6}& {\it Balaenoptera acutorostrata }& Cetacea & 25& 1 \\
\hline
0.9--1.0 & Giant anteater\cite[\#67-29]{www.brainmuseum.org}& {\it Myrmecophaga tridactyla }& Pilosa & 5.1& 1 \\
 & Tonkean macaque\cite[Fig.~7]{10.3389/fnana.2011.00029}& {\it Macaca tonkeana  }& Primates & 6.8& 2 \\
 & Lion\cite[\#62-79]{www.brainmuseum.org}& {\it Panthera leo }& Carnivora & 7.6& 1 \\
 & Yellow baboon\cite[Fig.~7]{10.3389/fnana.2011.00029}& {\it Papio cynocephalus  }& Primates & 8.0& 2 \\
 & Giant panda\cite[Fig.~3]{10.1002/cne.900840102}& {\it Ailuropoda melanoleuca }& Carnivora & 8.3& 1 \\
 & Western gorilla\cite[\#81-127]{www.brainmuseum.org}& {\it Gorilla gorilla }& Primates & 8.5& 1 \\
 & Common chimpanzee\cite[\#63-307]{www.brainmuseum.org}& {\it Pan troglodytes }& Primates & 8.6& 1 \\
 & Western lowland gorilla\cite[Fig.~1]{10.1002/ajp.20048}& {\it Gorilla gorilla gorilla }& Primates & 9.7& 1 \\
 & Polar bear\cite[\#62-250]{www.brainmuseum.org}& {\it Ursus maritimus }& Carnivora & 9.8& 1 \\
 & Hippopotamus\cite[Fig.~2]{10.1002/ar.23991}& {\it Hippopotamus amphibius }& Artiodactyla & 10& 1 \\
 & Bornean orangutan\cite[Fig.~2]{10.1002/ar.a.20256}& {\it Pongo pygmaeus  }& Primates & 11& 1 \\
 & California sea lion\cite[Fig.~10]{10.1002/cne.24188}& {\it Zalophus californianus }& Carnivora & 12& 1 \\
 & Human\cite[\#69-314]{www.brainmuseum.org}& {\it Homo sapiens }& Primates & 15& 2 \\
 & Humpback whale\cite[Fig.~1]{10.1007/s00429-014-0860-3}& {\it Megaptera novaeangliae }& Cetacea & 18& 1 \\
\hline
1.0--1.1 & Mountain gorilla\cite[Fig.~1]{10.1002/ajp.20048}& {\it Gorilla beringei beringei }& Primates & 9.6& 2 \\
 & Brown bear\cite[Fig.~1]{10.3389/fnana.2019.00079}& {\it Ursus arctos }& Carnivora & 9.8& 1 \\
 & Human\cite[Fig.~7]{10.3389/fnana.2011.00029}& {\it Homo sapiens }& Primates & 13& 2 \\
 & Common chimpanzee\cite[Fig.~1]{10.3389/fnana.2020.00055}& {\it Pan troglodytes }& Primates & 14& 1 \\
\hline
1.1--1.2 & Brown bear\cite[Fig.~7]{10.3389/fnana.2011.00029}& {\it Ursus arctos  }& Carnivora & 11& 1 \\
 & Humpback whale\cite[Fig.~2]{10.1002/ar.20407}& {\it Megaptera novaeangliae }& Cetacea & 26& 1 \\
\hline
1.2--1.3 & Asian elephant\cite[Fig.~5]{10.1016/j.brainresbull.2006.03.016}& {\it Elephas maximus }& Proboscidea & 23& 1 \\
 & Asian elephant\cite[Fig.~7]{10.1016/j.brainresbull.2006.03.016}& {\it Elephas maximus }& Proboscidea & 24& 1 \\
\hline
1.4--1.5 & Bowhead whale\cite[Fig.~1]{10.1002/ar.23991}& {\it Balaena mysticetus }& Cetacea & 25& 1 \\
\hline\end{tabular}
}
\end{table*}

\begin{table*}[htbp]
\caption{
Lissencephalic brains. Their hemisphere sizes are measured as illustrated in Fig.~\ref{Lis} and are plotted in Fig.~\ref{colored}. 
\label{smooth}}
\centering
\resizebox{\textwidth}{!}{
\begin{tabular}{c|lllc}
\hline
\textbf{Hemisphere size (cm)} & \textbf{Common name} & \textbf{Species name} & \textbf{Order} & \textbf{Brain size (cm)} \\
\hline
\hline
0.3--0.4 & Lesser horseshoe bat\cite[\#62-127]{www.brainmuseum.org}& {\it Rhinolophus hipposideros }& Chiroptera & 0.68 \\
\hline
0.5--0.6 & House mouse\cite[\#59-387]{www.brainmuseum.org}& {\it Mus musculus }& Rodentia & 1.0 \\
 & Merriam's kangaroo rat\cite[\#70-238]{www.brainmuseum.org}& {\it Dipodomys merriami }& Rodentia & 1.1 \\
\hline
0.6--0.7 & Common shrew\cite[\#64-25]{www.brainmuseum.org}& {\it Sorex araneus }& Eulipotyphla & 1.2 \\
 & North American deermouse\cite[\#59-390]{www.brainmuseum.org}& {\it Peromyscus maniculatus gracilis }& Rodentia & 1.3 \\
 & Eastern mole\cite[\#59-267]{www.brainmuseum.org}& {\it Scalopus aquaticus }& Eulipotyphla & 1.4 \\
\hline
0.7--0.8 & Mongolian gerbil\cite[\#60-352]{www.brainmuseum.org}& {\it Meriones unguiculatus }& Rodentia & 1.5 \\
 & Sugar glider\cite[\#64-20]{www.brainmuseum.org}& {\it Petaurus breviceps }& Diprotodontia & 1.5 \\
 & Degu\cite[\#66-40]{www.brainmuseum.org}& {\it Octodon degus }& Rodentia & 1.6 \\
 & Brown rat (Norway rat)\cite[\#63-463]{www.brainmuseum.org}& {\it Rattus norvegicus }& Rodentia & 1.6 \\
\hline
0.8--0.9 & Edible dormouse\cite[\#61-614]{www.brainmuseum.org}& {\it Glis glis }& Rodentia & 1.6 \\
 & Southern flying squirrel\cite[\#60-92]{www.brainmuseum.org}& {\it Glaucomys volans }& Rodentia & 1.7 \\
\hline
0.9--1.0 & Thirteen-lined ground squirrel\cite[\#67-132]{www.brainmuseum.org}& {\it Ictidomys tridecemlineatus }& Rodentia & 1.8 \\
 & Northern treeshrew\cite[\#62-279]{www.brainmuseum.org}& {\it Tupaia glis }& Scandentia & 1.8 \\
 & Greater glider\cite[\#64-37]{www.brainmuseum.org}& {\it Petauroides volans }& Diprotodontia & 2.0 \\
 & Tailless tenrec\cite[\#70-89]{www.brainmuseum.org}& {\it Tenrec ecaudatus }& Afrosoricida & 2.0 \\
 & Common ringtail possum\cite[\#64-19]{www.brainmuseum.org}& {\it Pseudocheirus peregrinus }& Diprotodontia & 2.0 \\
 & Squirrel glider\cite[\#64-28]{www.brainmuseum.org}& {\it Petaurus norfolcensis }& Diprotodontia & 2.0 \\
\hline
1.0--1.1 & Muskrat\cite[\#63-112]{www.brainmuseum.org}& {\it Ondatra zibethicus }& Rodentia & 2.1 \\
 & Bennett's chinchilla rat\cite[\#66-39]{www.brainmuseum.org}& {\it Abrocoma bennettii }& Rodentia & 2.1 \\
\hline
1.1--1.2 & Pygmy anteater\cite[Fig.~6]{10.1111/ahe.12501}& {\it Cyclopes didactylus }& Pilosa & 2.3 \\
 & Philippine tarsier\cite[\#61-193]{www.brainmuseum.org}& {\it Carlito syrichta }& Primates & 2.3 \\
\hline
1.2--1.3 & Long-tailed chinchilla\cite[\#65-103]{www.brainmuseum.org}& {\it Chinchilla lanigera }& Rodentia & 2.4 \\
 & Nine-banded armadillo\cite[Fig.~5]{10.1111/ahe.12501}& {\it Dasypus novemcinctus }& Cingulata & 2.4 \\
\hline
1.3--1.4 & Mountain beaver\cite[\#64-103]{www.brainmuseum.org}& {\it Aplodontia rufa }& Rodentia & 2.6 \\
 & Eastern gray squirrel\cite[\#60-144]{www.brainmuseum.org}& {\it Sciurus carolinensis }& Rodentia & 2.6 \\
 & Common brushtail possum\cite[\#64-29]{www.brainmuseum.org}& {\it Trichosurus vulpecula }& Diprotodontia & 2.8 \\
\hline
1.4--1.5 & European (domestic) rabbit\cite[\#73-211]{www.brainmuseum.org}& {\it Oryctolagus cuniculus }& Lagomorpha & 2.9 \\
 & Woodchuck (groundhog)\cite[\#61-770]{www.brainmuseum.org}& {\it Marmota monax }& Rodentia & 3.0 \\
\hline\end{tabular}
}
\end{table*}

\begin{table*}[htbp]
\caption{
Quasi-gyrencephalic brains with underdeveloped sulci. Their drawn sizes are measured as illustrated in Fig.~\ref{quasi} and Fig.~\ref{hint} and are plotted in Fig.~\ref{colored}.
\label{tablehint}}
\centering
\begin{tabular}{c|lllc}
\hline
\textbf{Drawn size (cm)} & \textbf{Common name} & \textbf{Species name} & \textbf{Order} & \textbf{Brain size (cm)} \\
\hline
\hline
0.5--0.6 & Senegal bushbaby\cite[\#61-686]{www.brainmuseum.org}& {\it Galago senegalensis }& Primates & 1.9 \\
 & Long-nosed potoroo\cite[\#65-55]{www.brainmuseum.org}& {\it Potorous tridactylus }& Diprotodontia & 2.4 \\
\hline
0.6--0.7 & Guinea pig\cite[Fig.~2]{10.1159/000123814}& {\it Cavia porcellus }& Rodentia & 2.0 \\
 & Guinea pig\cite[\#60-1]{www.brainmuseum.org}& {\it Cavia porcellus }& Rodentia & 2.0 \\
\hline
0.7--0.8 & Common marmoset\cite[\#54-64]{www.brainmuseum.org}& {\it Callithrix jacchus }& Primates & 2.5 \\
 & Tammar wallaby\cite[\#64-33]{www.brainmuseum.org}& {\it Macropus eugenii }& Diprotodontia & 2.9 \\
\hline
0.8--0.9 & Indian flying fox\cite[\#61-617]{www.brainmuseum.org}& {\it Pteropus giganteus }& Chiroptera & 2.5 \\
\hline
1.6--1.7 & Dugong\cite[Fig.~3]{10.1111/j.1439-0469.1985.tb00577.x}& {\it Dugong dugon }& Sirenia & 6.6 \\
 & Florida manatee\cite[\#85-32]{www.brainmuseum.org}& {\it Trichechus manatus latirostris }& Sirenia & 8.3 \\
\hline
1.9--2.0 & Florida manatee\cite[Fig.~20]{10.1017/S1464793106007019}& {\it Trichechus manatus latirostris }& Sirenia & 8.4 \\
\hline\end{tabular}
\end{table*}

\begin{table*}[htbp]
\caption{
Brains featuring shallow dimples or sulci, as shown in Fig.~\ref{dimple_irregular}. Generally larger than standard lissencephalic brains, these brains are often linked to lower-risk lifestyles compared to the quasi-gyrencephalic category, except for dugongs and manatees.
\label{dimple}}
\centering
\begin{tabular}{lllc}
\hline
 \textbf{Common name} & \textbf{Species name} & \textbf{Order} & \textbf{Brain size (cm)} \\
\hline
West European hedgehog\cite[\#61-559]{www.brainmuseum.org}& {\it Erinaceus europaeus }& Eulipotyphla & 2.0 \\
Snowshoe hare\cite[\#64-120]{www.brainmuseum.org}& {\it Lepus americanus }& Lagomorpha & 2.9 \\
North American porcupine\cite[\#62-541]{www.brainmuseum.org}& {\it Erethizon dorsatum }& Rodentia & 3.8 \\
North American beaver\cite[\#63-168]{www.brainmuseum.org}& {\it Castor canadensis }& Rodentia & 3.9 \\
North American beaver\cite[Fig.~4]{10.1159/000123818}& {\it Castor canadensis }& Rodentia & 4.8 \\
\hline\end{tabular}
\end{table*}

\begin{table*}[htbp]
\caption{
Brains with irregular gyrification patterns, as illustrated in Fig.~\ref{dimple_irregular}. These brains are generally larger than those in the quasi-gyrencephalic category, with the exception of dugongs and manatees, yet smaller than standard gyrencephalic brains. They typically feature a limited number of sulci with varying sizes.
\label{irregular}}
\centering
\begin{tabular}{lllc}
\hline
 \textbf{Common name} & \textbf{Species name} & \textbf{Order} & \textbf{Brain size (cm)} \\
\hline
Nine-banded armadillo\cite[\#60-465]{www.brainmuseum.org}& {\it Dasypus novemcinctus }& Cingulata & 2.9 \\
Three-striped night monkey\cite[\#69-255]{www.brainmuseum.org}& {\it Aotus trivirgatus }& Primates & 2.9 \\
Tasmanian pademelon\cite[\#65-59]{www.brainmuseum.org}& {\it Thylogale billardierii }& Diprotodontia & 3.3 \\
Common squirrel monkey\cite[\#61-672]{www.brainmuseum.org}& {\it Saimiri sciureus }& Primates & 3.4 \\
Mantled howler\cite[\#68-409]{www.brainmuseum.org}& {\it Alouatta palliata }& Primates & 4.3 \\
Western grey kangaroo\cite[\#62-127]{www.brainmuseum.org}& {\it Macropus fuliginosus }& Diprotodontia & 4.7 \\
Greater spot-nosed monkey\cite[\#63-490]{www.brainmuseum.org}& {\it Cercopithecus nictitans }& Primates & 5.4 \\
Collared mangabey\cite[\#63-489]{www.brainmuseum.org}& {\it Cercocebus torquatus }& Primates & 6.6 \\
\hline\end{tabular}
\end{table*}
\bibliography{sulci}

\end{document}